\documentclass[conference]{IEEEtran}
\IEEEoverridecommandlockouts
\usepackage{cite}
\usepackage{amsmath,amssymb,amsfonts}
\usepackage{algorithmic}
\usepackage{graphicx}
\usepackage{textcomp}
\usepackage{subfigure}

\usepackage{xcolor}
\def\BibTeX{{\rm B\kern-.05em{\sc i\kern-.025em b}\kern-.08em
    T\kern-.1667em\lower.7ex\hbox{E}\kern-.125emX}}
\begin{document}

\makeatletter 
\newcommand{\linebreakand}{%
  \end{@IEEEauthorhalign}
  \hfill\mbox{}\par
  \mbox{}\hfill\begin{@IEEEauthorhalign}
}
\makeatother 

\title{Statistical model for describing heart rate variability in normal rhythm and atrial fibrillation\\
\thanks{State task to USMU No. 730000F.99.1.BV10AA00006 ``Development of an integrated approach to personalized diagnosis, therapy and prevention of supraventricular cardiac arrhythmias of autonomic genesis''}
}

\author{

\IEEEauthorblockN{Nikita Markov$^{1,2}$}
\IEEEauthorblockA{\textit{$^{1}$Ural State Medical University} \\
\textit{$^{2}$Ural Federal University} \\
Yekaterinburg, Russia \\
shatzkarts@gmail.com}

\and

\IEEEauthorblockN{Ilya Kotov$^{2}$}
\IEEEauthorblockA{\textit{$^{2}$Ural Federal University} \\
Yekaterinburg, Russia \\
Ilya.Kotov@urfu.me}

\and

\IEEEauthorblockN{Konstantin Ushenin$^{1,2,3}$}
\IEEEauthorblockA{\textit{$^{1}$Ural State Medical University} \\
\textit{$^{2}$Ural Federal University} \\
\textit{$^{3}$Institute of Immunology and Physiology UrB RAS }\\
Yekaterinburg, Russia \\
konstantin.ushenin@urfu.ru}

\linebreakand

\IEEEauthorblockN{Yakov Bozhko$^{1}$}
\IEEEauthorblockA{\textit{$^{1}$Ural State Medical University} \\
Yekaterinburg, Russia \\
yakov-bozhko@yandex.ru}
}

\maketitle

\begin{abstract}
Heart rate variability (HRV) indices describe properties of interbeat intervals in electrocardiogram (ECG). Usually HRV is measured exclusively in normal sinus rhythm (NSR) excluding any form of paroxysmal rhythm. Atrial fibrillation (AF) is the most widespread cardiac arrhythmia in human population. Usually such abnormal rhythm is not analyzed and assumed to be chaotic and unpredictable. Nonetheless, ranges of HRV indices differ between patients with AF, yet physiological characteristics which influence them are poorly understood. In this study, we propose a statistical model that describes relationship between HRV indices in NSR and AF. The model is based on Mahalanobis distance, the k-Nearest neighbour approach and multivariate normal distribution framework. Verification of the method was performed using 10 min intervals of NSR and AF that were extracted from long-term Holter ECGs. For validation we used Bhattacharyya distance and Kolmogorov-Smirnov 2-sample test in a k-fold procedure. The model is able to predict at least 7 HRV indices with high precision.
\end{abstract}


\section{Introduction}

Atrial fibrillation (AF) is one of the most widespread in population heart disease. AF drastically impacts health condition of suffering patients. Also, this disease generates significant financial burden to national healthcare systems.

Heart rate variability (HRV) in paroxysmal or persistent AF is not well studied, and some studies describe HRV in AF as chaotic and unpredictable \cite{garfinkel1997quasiperiodicity, van1993heart, khan2021heart}. However, without any doubts, ECG of atrial fibrillation is unique for each patient, and ranges of HRV indices also differ between patients with AF. HRV of the normal rhythms depends on age, sex, activity of sympathetic and parasympathetic nervous systems \cite{ishaque2021trends, shaffer2017overview}. The same physiological characteristics influence cardiac rhythm in AF, but these relationships are poorly studied. 

In this preliminary study, we propose a statistical model that connects HRV indices measured in normal cardiac rhythm and HRV indices measured in paroxysms of AF. The model is based on an assumption that HRV indices calculated in consecutive 10-minute ECG recordings of a single patient follow multivariate normal distribution (MND). The study describes the statistical model and method of its training on real data. Validation of the method and model performed using k-fold method on 30 cases.

\section{Methods}

\subsection{Mathematical framework}

The main assumption of our work is heart rate variability indices follow MND rules. Probability density function of MND is formulated as follows:
\begin{equation*}
f(\mathbf{x}) = (2\pi)^{-\frac{n}{2}}\big| \boldsymbol{\Sigma} \big|^{-\frac{1}{2}} \exp(-(\mathbf{x}-\boldsymbol{\mu})^{T} \boldsymbol{\Sigma}^{-1} (\mathbf{x}-\boldsymbol{\mu}) /2).
\end{equation*}
Thus, MND $\mathcal{N}(\boldsymbol{\mu}, \boldsymbol{\Sigma})$ is defined by a mean vector $\boldsymbol{\mu} = (\mu_1\ \mu_2 ..\ \mu_n)^T$ and covariance matrix $\boldsymbol{\Sigma}$, where $n$ is dimension of the vector space.
Mahalanobis distance is a measure of distance defined between distribution $\boldsymbol{L} \sim \mathcal{N}(\boldsymbol{\mu}, \boldsymbol{\Sigma})$ and a vector $\mathbf{z} = (z_1\ z_2\ ..\ z_n)^T$:
\begin{equation*}
D_{M}(\boldsymbol{L},\mathbf{z})=\sqrt{(\mathbf{z}-\boldsymbol{\mu})^T \boldsymbol{\Sigma} (\mathbf{z}-\boldsymbol{\mu})}.
\end{equation*}

$X_{p}$ is a data matrix that presents sampling from a random vector $\mathbf{X}_{p}$:
\begin{equation*}
X_p = \begin{pmatrix}
x_{p,1,1} & x_{p,1,2} & .. & x_{p,1,I}\\
x_{p,2,1} & x_{p,2,2} & .. & x_{p,2,I}\\
.. & .. & .. & ..\\
x_{p,S_p,1} & x_{p,S_p,2} & .. & x_{p,S_p,I}
\end{pmatrix}.
\end{equation*}
Here, variable $x_{p,s,i}$ is a value of HRV coefficient with number $i \in [1,2,..,I]$ measured in a short ECG signal $s \in [1,2,..,S_p]$; $p \in [1,2,..,P]$ is an index of a patient $\{X_p\}_{p=1}^{P}$. Note, the number of samples $S_p$ depends on a patient and is not fixed.

The method proposed in our work is based on assumption that HRV indices follow normal distribution rules: $\forall p, \boldsymbol{X}_{p} \sim \mathcal{N}(\boldsymbol{\mu}_p, \boldsymbol{\Sigma}_p)$ where $\boldsymbol{\mu}_p$ is a centroid and $\boldsymbol{\Sigma}_p$ is a covariance matrix.

Training dataset consists of pairs $(X_p,Y_p)_{p=1}^{P}$, where $\mathbf{X}_{p} \sim \mathcal{N}(\boldsymbol{\mu}_p, \boldsymbol{\Sigma}_p)$ and $\mathbf{Y}_{p} \sim \mathcal{N}(\boldsymbol{\xi}_p, \boldsymbol{\Psi}_p)$. Series of data matrices $X_p$ present data for normal sinus rhythm. Series of data matrices $Y_p$ present data for atrial fibrillation. Trained algorithm processes signals of a new patient $\{\tilde{X}_{\tilde{p}}\}_{\tilde{p}=1}^{\tilde{P}}$ that was not used on the training. 

Model training is a construction of linear transformations $\hat{Y_p} = A_p(X_p)$ for pairs of observations $(X_p,Y_p)_{p=1}^{P}$. The transformation optimizes a statistical distance $D(Y_p, \hat{Y}_p) \rightarrow \min$ between sampling of $\mathbf{Y}_p \sim \mathcal{N}(\boldsymbol{\xi}_p, \boldsymbol{\Psi}_p)$ and the predicted one $\hat{\mathbf{Y}}_{p} \sim  \mathcal{N}(\hat{\boldsymbol{\xi}}_p, \hat{\boldsymbol{\Psi}}_p)$. This minimization problem is defined independently for each patient: 
\begin{align*}
\|E({\mathbf{Y}_{p}}) - E(\hat{\mathbf{Y}}_{p})\|_{2} &\rightarrow \min, \\
\| \operatorname{cov}(\mathbf{Y}_{p}) - \operatorname{cov}(\hat{\mathbf{Y}}_{p}) \|_F &\rightarrow \min,
\end{align*}
where $\| \cdot \|_2$ is Euclidean norm, $\| \cdot \|_F$ is Frobenius matrix norm, $E(\cdot)$ is a mean vector, and $\operatorname{cov}(\cdot)$ is a covariance matrix.

We define $A_p(\mathbf{X})$ as a combination of shift, rotation and scaling operators. The shift operation is defined as subtraction of the mean vectors for original and the target sampling set. Rotation operator is required to transform the original sampling set to a form where its covariance matrix is diagonal. We obtain it as an inverse of a matrix of eigenvectors of $cov(\mathbf{X}_p)$ which align principal directions of the sampling set to linear space basis \cite{abdi2010principal}. Diagonal covariance matrix indicates no pairwise correlations between variables which means they can be freely scaled. Following the general rule $\operatorname{cov}(\alpha \mathbf{X}_p) = \alpha^2 \operatorname{cov}(\mathbf{X}_p)$ we can scale the intermediate distribution so values of its diagonal covariance matrix equal to eigenvalues of $\Psi_p$. Thus, the scaling operator is a diagonal matrix $\Lambda_p = diag(\sqrt{\lambda_1^{X p} \cdot \lambda_1^{Y p}},.., \sqrt{\lambda_I^{X p} \cdot \lambda_I^{Y p}})$, where $\{\lambda_i^{X p}\}_{i=1}^I$ and $\{\lambda_i^{Y p}\}_{i=1}^I$ are eigenvalues of $cov(\mathbf{X}_p)$ and $cov(\mathbf{Y}_p)$ respectively \cite{abdi2010principal}. The last step involves another general rotation to align the distribution with principal directions of $\mathcal{N}(\boldsymbol{\xi}_p, \boldsymbol{\Psi}_p)$, using a matrix of eigenvectors of $cov(\mathbf{Y}_p)$ as operator.

The transformation $A_p(\boldsymbol{\theta}): R^{s \times i} \rightarrow R^{s \times i}$ can be formulated as follows:
\begin{equation*}
\label{linear}
    A_p(\mathbf{X}) = (\mathbf{X}^T - E(X_p)) \cdot V({X_p})^{-1} \cdot \Lambda_p \cdot V({Y_p}) + E(Y_p).
\end{equation*}
Here $V(\cdot)$ is a rotation matrix where rows are eigenvectors of respective covariance matrix. Fig. \ref{fig:rotplot} showcases each step of the transformation in case of arbitrary bi-variate normal distribution samples.

\begin{figure}[t]
    \centering
    \subfigure
    {
        \includegraphics[width=3in]{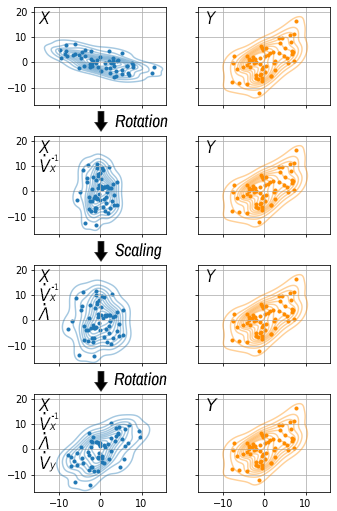}
        \label{fig:kotplot}
    }
    \caption{Transformation example showcasing the work of rotation and scaling operators.}
    \label{fig:rotplot}
\end{figure}

Several transformations obtained from the training dataset are joined together using the k-Nearest Neighbour approach. The method uses Mahalanobis distance and the closest neighbours. The shortest distance defined using the softmax function. The joined transformation take vector of indices for the one sample of NSR HRV vector $\tilde{\mathbf{x}}$, and returns a single sample of AF HRV vector $\tilde{\mathbf{y}}$:

\begin{align*}
    w_p &= D_{M}(\mathbf{X}_p, \tilde{\mathbf{x}}),\\
    \tilde{\mathbf{y}} &= \sum_{p=1}^{P} \frac{\exp(w_p)}{\sum_{p=1}^{P} \exp(w_p)} A_p(\tilde{\mathbf{x}}).
\end{align*}

\subsection{Model verification}

Training and verification of the statistical model were performed on two databases from the open  databank of physiological signals PhysioNet \cite{goldberger2000physiobank}. Long Term Fibrillation Database (LTAFDB) \cite{petrutiu2007abrupt} consists of data from 84 patients with 84 ECG recordings of duration between 24 and 25 hours. Atrial Fibrillation Database (AFDB) \cite{moody1983new} is made up of 23 patients, each represented as 10-hour ECG recording. Both databases are supplied with annotation markers for NSR and AF rhythms as well as premature ectopic beats.

Each recording from both databases was divided into 10-minute segments with 5-minute window overlaps. Segments were labeled as NSR and AF according to annotations. In the case an ECG segment included both NSR and AF episodes it was excluded from analysis. Likewise recordings with less than 15 segments ($S=15$) per rhythm type were withdrawn entirely ensuring that each analyzed recording included at least 80 minutes of NSR and AF episodes. Recordings of 30 patients were used for model training and the verification ($P=30$).

A feature vector of HRV indices ($X_{p,s}$) was calculated for each ECG segment $s$ using interbeat intervals with excluded ectopic beats (according to database annotations). Each segment was represented by 6 time-domain indices (RMSSD, MeanNN, SDNN, IQRNN, pNN50, pNN20) \cite{shaffer2017overview}, 2 geometric indices (TINN, HTI) \cite{heart1996standards}, 3 frequency domain indices (LF, HF, VHF) \cite{shaffer2017overview}, 2 nonlinear indices (SD1, SD2) \cite{shaffer2017overview}, 2 indices of heart rate fragmentation (PIP, PAS) \cite{costa2017heart}, 2 indices of rhythm asymmetry (AI, PI) \cite{yan2017area} and entropy index ApEn \cite{shaffer2017overview}. Thus each HRV feature vector included 18 values ($I=18$) and each ECG recording was represented by a pair of HRV matrices of NSR and AF rhythms. We assume them to be samples of 18-dimensional multivariate normal distribution ($n=18$). Thus, our whole population data can be represented as a set of matrix pairs $(X_p, Y_p)_{p=1}^{30}$.

Model validation were performed using k-fold method \cite{refaeilzadeh2009cross} of cross-validation with $k=5$. Denote the training dataset as $(X_p, Y_p)_{p=1}^{P}$ with $P=24$, and the test dataset as $(\tilde{X}_{\tilde{p}}, \tilde{Y}_{\tilde{p}})_{\tilde{p}=1}^{\tilde{P}}$ with $\tilde{P}=6$.

The minimization problem converges down to zero. However, predicted values of the method output are far from the expected values in some folds or indices. For this reason, we applied two method which compare distributions of the predicted values.

With the assumption that predicted data follows normal distribution rules we use Bhattacharyya distance \cite{bhattacharyya1946measure} to measure the similarity between expected sample set $\mathbf{Y} \sim {\mathcal  {N}}({\boldsymbol  \mu }_{i},\,{\boldsymbol  \Sigma }_{i})$ and predicted sample set $\tilde{\mathbf{Y}} \sim {\mathcal  {N}}({\tilde{\boldsymbol{\mu}} }_{i},\,\tilde{\boldsymbol{\Sigma}}_{i})$. 

\begin{align*}
D_{B} =& {1 \over 8}(\boldsymbol{\mu}-\tilde{\boldsymbol{\mu}})^{T} \hat{\boldsymbol{\Sigma}}^{{-1}}(\boldsymbol{\mu}-\tilde{\boldsymbol{\mu}})+\\
 &+{1 \over 2}\ln \,\left({\det \hat{\boldsymbol{\Sigma}} \over {\sqrt  {\det \boldsymbol{\Sigma}\,\det \tilde{\boldsymbol{\Sigma}}}}}\right),\\
\hat{\boldsymbol{\Sigma}} =& \frac{\boldsymbol{\Sigma}+\tilde{\boldsymbol{\Sigma}}}{2}.
\end{align*}

Additionally, distribution of the each predicted HRV index $\tilde{y}_{p,s,i}$ were compared with expected $y_{p,s,i}$ using two-sample Kolmogorov–Smirnov test \cite{berger2014kolmogorov}:

\begin{align*}
D_{n,m} =& \sup _{x}|F_{1,n}(x)-F_{2,m}(x)|,\\
D_{n,m} >& c(\alpha ){\sqrt {\frac {n+m}{n\cdot m}}},\\
c\left(\alpha \right) =& {\sqrt {-\ln \left({\tfrac {\alpha }{2}}\right)\cdot {\tfrac {1}{2}}}},
\end{align*}
where $n$ and $m$ are the sizes of first and second sample respectively, $\alpha$ is level of statistical significance, $c(\alpha)$ is Kolmogorov–Smirnov statistics, and  $F_{1,n}$ and $F_{2,m}$ are the empirical distribution functions of the first and the second samples respectively.

\section{Results}
We consider a statistical model that predicts a sample of HRV indices in AF using HRV samples measured in 10-minute NSR ECG segments. The model uses population data from patients with paroxysmal AF.

\begin{table}
\caption{5-fold cross-validation results. Similarity of samples measured with Bhattacharyya distance for each data split.}\label{tbl:kfold}
\centering
\begin{tabular}{l|ccc}
Split & Mean Bhatt, & Best Bhatt. & Worst Bhatt. \\
\hline
1 & 10.78  &  5.94 & 16.26 \\
2 & 12.36 & 8.27 & 17.08 \\
3 & 7.89 & 5.06 & 11.88 \\ 
4 & 14.83 & 3.97 & 39.27 \\
5 & 11.44 &  6.96 & 17.34 \\
\hline
All & 11.46 & 3.97 & 39.27 \\
\end{tabular}

\end{table}

\begin{table}
\caption{5-fold cross-validation results of goodness of fit per HRV index. The highest p-value is chosen for each split.}\label{tbl:ksp}
\centering
\begin{tabular}{l|ccccc|c}
Index & Split 1 & Split 2 & Split 3 & Split 4 & Split 5 & Passing \\
& \textit{p-val.} & \textit{p-val.} & \textit{p-val.} & \textit{p-val.} & \textit{p-val.} & splits \\
\hline
RMSSD  & 0.073 & 0.129 & \textbf{0.002}  & 0.051 & \textbf{0.012} & 2 \\ 
MeanNN  & \textbf{0.001} & \textbf{0.001} & \textbf{0.007}   & \textbf{0.000} &  \textbf{0.020} & \textbf{5} \\ 
SDNN  & 0.101 & 0.093 & \textbf{0.002}  & 0.066 & \textbf{0.003} & 2 \\ 
IQRNN  & \textbf{0.032} & 0.075 & \textbf{0.013}  & 0.165 & \textbf{0.016} &  3 \\ 
pNN50  & \textbf{0.002} & \textbf{0.001} & \textbf{0.000}  & \textbf{0.008} & \textbf{0.020} & \textbf{5} \\
pNN20  & \textbf{0.001} & \textbf{0.001} & \textbf{0.000}  & \textbf{0.007} & \textbf{0.007} & \textbf{5} \\
TINN  & \textbf{0.047} & \textbf{0.000} & 0.064 & \textbf{0.008}  & 0.074 & 3 \\
HTI  & \textbf{0.000} & 0.078 & \textbf{0.002}  & \textbf{0.005} & 0.127 & 3 \\ 
LF  & \textbf{0.000} & \textbf {0.020} & \textbf{0.020}  & \textbf{0.010} & 0.168 & 4 \\
HF  & \textbf{0.023} & \textbf{0.046} & \textbf{0.031}  & \textbf{0.032} & 0.068 & 4 \\
VHF  & \textbf{0.000} & 0.075 & \textbf{0.021}  & 0.067 & \textbf{0.003} & 3 \\
SD1  & 0.073 & 0.129 &  \textbf{0.002}  & 0.051 & \textbf{0.012} & 2 \\
SD2  & 0.114 & \textbf{0.041} & 0.074  & 0.09 & 0.051 & 1 \\
PIP  & \textbf{0.043} & 0.134 & \textbf{0.007}  & 0.081 & \textbf{0.007} & 3 \\
PAS  & \textbf{0.044} & 0.055 & \textbf{0.003}  & 0.091 & \textbf{0.001} & 3 \\
AI  & \textbf{0.013} & \textbf{0.004} & 0.206  & \textbf{0.001} & \textbf{0.015} & 4 \\
PI  & 0.051 & \textbf{0.010} & 0.243  & 0.169 & 0.261 & 1 \\
ApEn  & \textbf{0.001} & \textbf{0.006} & \textbf{0.001}  & 0.137 & \textbf{0.013} & 4 \\
\hline
Passing  & 13 & 10 & 14 & 8 & 12 \\
indices

\end{tabular}

\end{table}

To verify the model we represented 30 ECG recordings with long episodes of paroxysmal atrial fibrillation as point distributions of NSR and AF HRV indices. A 5-fold validation was utilized for model verification so 25 and 5 recordings made up training and testing population for each of 5 splits. Table \ref{tbl:kfold} presents similarity measurements with Bhattacharyya distance between predicted and test AF HRV point distributions for each split. Table \ref{tbl:ksp} presents goodness of fit results for each individual index measured with Kolmogorov-Smirnov 2-sample test. 


The number of indices that pass goodness of fit test (the highest test p-value $< 0.5$) correlates with the mean Bhattacharyya distance for each split. The best split with 7.89 mean distance shows reliable prediction for 14 indices out of 18, but the worst split with 14.83 mean distance indicates reliable prediction for only 8 indices out of 18. Certain HRV metrics are predicted regardless of the data split. MeanNN (mean of interbeat intervals), pNN50 / pNN20 (percent of adjacent RR-intervals that differ by 50/20 ms) pass goodness of fit test in each split. Indices LF (low frequency power), HF (high frequency power), AI (poincare plot area index) and ApEn (approximate entropy) pass the test for 4 splits out of 5. Some of these indices consistently show the clearest distinction between NSR and AF rhythms \cite{ishaque2021trends} which may explain reliability of the fit.

\section{Discussion and Conclusion}

This study presents the statistical model and includes the method of its training on the real data. The model uses data of HRV indices calculated on normal rhythm and predicts HRV indices calculated on paroxysmal rhythm. 

Validation of the model was performed on the set of ECG intervals obtained from 30 patients. Validation results show that HRV in AF is not entirely chaotic, and may be partially explained by HRV in NSR. There are seven HRV indices (MeanNN, pNN50, pNN20, LF, HF, ApEn, AI) that may be reliably predicted in AF using HRV indices of the normal rhythm.

We suppose that our model is the first step towards new class of methods for HRV processing. The future development of the presented method may help in prediction of HRV attributes of long-term ECG records (12-24 hours) using HRV measured on short on short ECG recording. Likewise, the mathematical framework applied for the model construction is suitable for further studies of HRV in anomalous types of cardiac rhythm.

\bibliographystyle{./IEEEtran}
\bibliography{bibliography.bib}

\end{document}